# Engineering Psychological Safety in Autonomous Vehicles

## A Systems-Theoretic Framework for Psychological Safety in Autonomous Vehicles and its Validation in Real-World Scenarios

Yandika Sirgabsou[a], Benjamin Hardin[b], François Leblanc[a] *,
Efi Raili[c], David Jackson[c], Pericle Salvini[b], Lars Kunze[d], Marina Jirotka[b]
[a]*Capgemini Engineering, Toulouse, France;*
[b]*Dept. of Computer Science, University of Oxford, UK;*
[c]*Capgemini Engineering, UK;*
[d]*Bristol Robotics Laboratory, University of the West of England, UK;*

*François Leblanc @ francois.a.leblanc@capgemini.com ; Capgemini Engineering, 4 Av. Didier Daurat, 31700 Blagnac, France

**ABSTRACT**

Despite rapid technological advances, the societal acceptability of autonomous vehicles (AVs) remains limited by psychological barriers that extend beyond traditional concerns of physical safety. While factors such as trust and perceived safety are known to influence user acceptance, there is a lack of formalized mechanisms and engineering methods to systematically identify, assess, and mitigate psychological risks arising from human-AV interactions.

To address this gap, this work proposes and validates a systems-theoretic framework for the assessment of psychological safety in autonomous vehicles. First, a comprehensive psychological safety risk model is defined, extending the Systems-Theoretic Accident Model and Processes (STAMP) to incorporate key psychological constructs such as trust, perceived control, predictability, and perceived support. Based on this model, a hazard analysis method (AV-PsySafe) is developed to systematically identify psychological hazards, unsafe control actions, and loss scenarios, while introducing a Psychological Safety Integrity Level (PsySIL) to support risk prioritization.

Second, the applicability and relevance of the framework are evaluated through its deployment in realistic autonomous vehicle scenarios. A structured validation approach is implemented, including a methodological guide, standardized analysis templates, and the collection of analyst feedback. The results demonstrate that the framework can be consistently applied by practitioners, producing meaningful insights into psychological risks.

Overall, this work establishes both the theoretical foundations and practical feasibility of a unified approach to co-assessing psychological and physical safety in autonomous systems, contributing to more human-centred and trustworthy AV development.

## 1. INTRODUCTION

The advent of autonomous vehicles (AVs) represents a major transformation in transportation, with the promise of significantly improving road safety, efficiency, and accessibility. However, despite continuous technological progress, the widespread adoption of AVs remains constrained by persistent societal concerns (Shariff et al., 2017; Yoo et al., 2025). While traditional safety engineering has primarily focused on mitigating physical risks, emerging evidence highlights that psychological factors play a critical role in shaping user acceptance and interaction with AVs (Cugurullo & Acheampong, 2024; Prasetio & Nurliyana, 2023; Shariff et al., 2021).

In particular, constructs such as trust, perceived safety, perceived control, and predictability have been identified as key determinants of users' willingness to adopt and rely on autonomous driving systems (Liu et al., 2019; Meyer-Waarden & Cloarec, 2022; Xu et al., 2018). However, although these psychological aspects have been extensively studied in behavioural and human factors research, they remain insufficiently addressed within system safety engineering practices. Current standards and methodologies largely emphasize physical harm prevention and lack formal mechanisms to capture how system behaviour and human-automation interaction can give rise to psychological hazards.

This gap is especially critical given that psychological risks may emerge even in the absence of physical danger. For example, unexpected vehicle behaviour, lack of transparency, or inadequate feedback may lead to anxiety, loss of trust, or reduced perceived control, ultimately undermining both user experience and system acceptability (Brell et al., 2019; Kenesei et al., 2022; Li et al., 2019; Thomas et al., 2020). Consequently, there is a growing need for engineering frameworks that treat psychological safety not as a secondary or subjective concern, but as an integral component of overall system safety.

To address this challenge, prior work introduced AV-PsySafe, a systems-theoretic framework for psychological safety in autonomous vehicles, based on STAMP (Systems-Theoretic Accident Model and Processes) and STPA (Systems-Theoretic Process Analysis), its associated hazard analysis method (Sirgabsou et al., 2025). This framework provides a conceptualization of psychological safety through a dedicated risk model, defining psychological losses, hazards, and safety goals, as well as a structured method (AV-PsySafe) to identify how control actions and interaction mechanisms within the system may generate psychological risk. However, this initial contribution was primarily theoretical and demonstrated on limited, illustrative use cases. As a result, two key concerns remained open:

- How can psychological risk be systematically assessed and quantified in practice?
- To what extent can the framework be applied by independent analysts in realistic engineering contexts?

These two concerns motivate the present work and are subsequently translated into four research questions addressing the framework's applicability, utility, industrial relevance, and methodological robustness, as detailed later in section 4.1.

To address these concerns, the paper proposes an enhancement and a validation of the AV-PsySafe framework. While previous work established the conceptual foundations of psychological safety assessment for autonomous vehicles, the present contribution focuses on making the AV-PsySafe framework operational through a realistic application in an engineering context (Sirgabsou et al., 2025).

First, the framework is enhanced through the refinement of its risk model, the introduction of structured methodological guidance, standardised analysis templates, refined PsySIL assessment support, and explicit traceability mechanisms linking psychological safety artefacts throughout the analysis process.

Second, the applicability, utility, industrial relevance, and robustness of the framework are evaluated through an independent validation study conducted within the Route 25 autonomous driving project (Route 25, n.d.). This project will be presented in section 4.2 for further details. The validation combines technical analysis results and structured analyst feedback as complementary sources of evidence.

The contributions of this paper are therefore twofold:

1. The transformation of AV-PsySafe from a conceptual framework into an engineering-ready methodology through the introduction of methodological, traceability, and risk assessment enhancements.
2. The validation of the resulting methodology through its application by an independent Validation Team in a realistic autonomous driving use case.

Through these contributions, the paper provides initial evidence that psychological safety can be systematically analysed and integrated within existing automotive safety engineering processes.

## 2. STATE OF THE ART

### 2.1. PSYCHOLOGICAL DETERMINANTS OF AUTONOMOUS VEHICLE ACCEPTANCE

Research on autonomous vehicles (AVs) has focused foremost on technological performance and physical safety. However, recent studies increasingly emphasize the role of psychological factors in shaping user acceptance and experience (Cugurullo & Acheampong, 2024; Prasetio & Nurliyana, 2023; Shariff et al., 2021). Specifically, trust, perceived safety, perceived control, and predictability have been identified as critical determinants influencing users' willingness to engage with autonomous driving systems (Hegner et al., 2019; Ma & Zhang, 2021; Zhang et al., 2021).

These psychological factors are closely linked to broader human–automation interaction principles, where user experiences, perceptions, and expectations strongly influence system effectiveness and acceptance (Ghazizadeh et al., 2011; Muir & Moray, 1996). Moreover, concerns related to ethical decision-making, uncertainty, and lack of transparency have been shown to create psychological barriers to AV adoption (Shariff et al., 2017; Yoo et al., 2025). All of these factors are encompassed under the more recent and broader term *psychological safety* of autonomous vehicles (BAILEY, 2026; Prasetio & Nurliyana, 2023; Sirgabsou et al., 2025).

Despite this growing body of knowledge, psychological factors are predominantly treated as outcomes of user studies rather than as engineering parameters that can be systematically modelled, analysed, and addressed during system design. The goal of this work is to enable incorporation of these psychological factors into proactive system design.

### 2.2. LIMITATIONS OF CURRENT SAFETY ENGINEERING APPROACHES

In contrast, safety engineering for AVs has been largely structured around physical risk mitigation, relying on well-established standards such as ISO 26262 (*ISO 21448:2022 - Road Vehicles — Safety of the Intended Functionality*, n.d.) and *ISO 21448* or SOTIF (*ISO 21448:2022 - Road Vehicles — Safety of the Intended Functionality*, n.d.). These approaches define safety in terms of the prevention of physical harm through the identification and mitigation of hazards, typically quantified using severity, exposure, and controllability metrics.

However, such frameworks are inherently limited when addressing psychological safety. They focus on objective and measurable failures, whereas psychological risks are often subjective, context-dependent, and emergent from human–automation interaction (Manger et al., 2025). Similarly, they do not explicitly account for non-physical losses, such as loss of trust, anxiety, or confusion. As a result, psychological considerations are generally addressed indirectly through interface design recommendations rather than formal safety processes.

This limitation is reinforced by classical human factors insights showing that inadequate feedback, poor transparency, or unexpected system behaviour can significantly affect user trust and perception, even when systems operate safely from an engineering standpoint (Lee & See, 2004; Norman, 1990). Consequently, existing safety frameworks do not fully capture the multidimensional nature of safety in autonomous systems.

### 2.3. SYSTEMS-THEORETIC APPROACHES AND EXTENSION TO PSYCHOLOGICAL SAFETY

The systems-theoretic approach to safety, through the Systems-Theoretic Accident Model and Processes (STAMP) and the Systems-Theoretic Process Analysis (STPA), offer a broader perspective by conceptualizing safety as an emergent property arising from interactions within complex systems (N. Leveson, 2004; N. G. Leveson, 2012; N. Leveson & Thomas, 2018). Unlike traditional safety approaches, STAMP enables the consideration of non-physical losses and interaction-based hazards, making it particularly suitable for analysing socio-technical systems such as AVs (Abdulkhaleq et al., 2017; Mahajan et al., 2017; Nouri et al., 2023)

Building on this perspective, recent work has proposed adopting systems-theoretic frameworks to incorporate psychological safety. In particular, the AV-PsySafe framework introduced a structured risk model for psychological safety based on STAMP and a corresponding hazard analysis method based on STPA (Sirgabsou et al., 2025). This frgoals andfines psychological losses, hazards, and safety goals, and provides a systematic approach to identifying within a dynamic model of the system how unsafe control actions may lead to psychological risks.

Furthermore, it introduces the concept of a Psychological Safety Integrity Level (PsySIL), enabling the rating of psychological risks in a manner analogous to traditional Automotive Safety Integrity levels (ASIL) (Sirgabsou et al., 2025). This represents a significant step toward the formalization and engineering integration of psychological safety.

### 2.4. Remaining gaps: validation and integration in engineering practice

Despite these advances, current contributions remain largely conceptual and exploratory. The proposed frameworks have primarily been demonstrated on simplified or illustrative use cases, which limits the assessment of their practical applicability in real-world engineering contexts.

In particular, several key challenges remain unaddressed. First, there is limited evidence regarding whether such frameworks can be effectively applied by practitioners, including their usability, learnability, and ability to produce consistent and meaningful analysis results. Second, there is a lack of structured guidance on how psychological safety analyses can be made operational within standard development processes, including traceability, documentation, and integration with existing safety artefacts.

Finally, although psychological safety is closely linked to physical safety in shaping overall system acceptability, there is still no established approach for their co-assessment within a unified safety engineering framework. This represents a critical gap, as interactions between physical system behaviour and user perception may significantly influence both safety outcomes and user trust (Sirgabsou et al., 2025).

## 3. The AV-PsySafe Framework

### 3.1. Overview of AV-PsySafe Framework

The AV-PsySafe framework is a systems-theoretic approach for assessing psychological safety risks arising from human interaction with autonomous vehicles (Sirgabsou et al., 2025). It extends STAMP and traditional safety concepts beyond physical harm by explicitly considering non-physical losses such as loss of trust, loss of perceived control, anxiety, confusion, and perceived lack of support. At a conceptual level, the framework consists of two complementary components: a *psychological safety risk model* that defines the concepts and relationships required to represent psychological risk, and *AV-PsySafe*, a hazard analysis method used to identify and analyse the mechanisms through which psychological losses may arise during human–automation interactions.

#### 3.1.1. Psychological Safety Risk Model

The AV-PsySafe framework defines psychological safety as the absence of unacceptable psychological losses arising from human–automation interaction (Sirgabsou et al., 2025). In addition, it introduces a structured risk model for autonomous vehicle psychological safety. The model structures the analysis through a set of interconnected concepts, including psychological stakes, psychological losses, psychological hazards, psychological safety goals, and psychological contexts encompassing system, environmental, and human factors.

The model also introduces the Psychological Safety Integrity Level (PsySIL), a risk prioritisation mechanism inspired by the Automotive Safety Integrity Level (ASIL) concept used in ISO 26262 functional safety assessments (*ISO 26262 [2018] Road Vehicles-Functional Safety-Part 1*, 2018).

Together, these elements provide a structured foundation for identifying, analysing, prioritising, and mitigating psychological risks in autonomous vehicle systems. However, while the original AV-PsySafe framework established these conceptual foundations, it remained primarily conceptual and provided limited methodological support for deployment within industrial engineering projects.

#### 3.1.2. Psy-STPA

AV-PsySafe also defines Psy-STPA, a psychological hazard analysis method adapted from STPA (N. G. Leveson, 2012; N. Leveson & Thomas, 2018). Psy-STPA systematically identifies how unsafe control actions and inadequate feedback within socio-technical control structures can give rise to psychological hazards and subsequent psychological losses (Sirgabsou et al., 2025).

The method follows a five-step sequential process:

- **Step 0 – Preparation:** system and boundary definition, scope specification, stakeholder identification, and psychological stakes identification;
- **Step 1 – Define the purpose of the analysis:** identification of psychological losses, hazards, and specification of psychological safety goals;
- **Step 2 – Model the psychological safety control structure:** modelling of human–autonomy–system control and feedback relationships based on STAMP control structures (Leveson, 2012);
- **Step 3 – Identify psychologically Unsafe Control Actions (UCAs):** identification of control actions that may contribute to psychological hazards under specific contexts;
- **Step 4 – Identify psychological loss scenarios:** analysis of causal pathways through which unsafe control actions may lead to psychological losses.

Together, the psychological safety risk model and Psy-STPA establish the conceptual and methodological foundations of AV-PsySafe. The risk model defines how psychological risks are represented and prioritised, while Psy-STPA provides a systematic process for their analysis. These foundations enable explicit traceability between psychological stakes, losses, hazards, safety goals, responsibilities, Unsafe Control Actions, and loss scenarios, supporting the integration of psychological safety considerations within systems engineering and safety engineering activities (Sirgabsou et al., 2025).

### 3.2. Framework Enhancements Introduced in this Work Prior to Validation

Building upon the original AV-PsySafe framework, this work introduces a set of enhancements aimed at supporting its practical application in engineering contexts. While the initial framework established the conceptual foundations of psychological safety assessment for autonomous vehicles, it provided limited support for deployment within industrial development activities. The enhancements introduced in this work therefore focus on improving methodological guidance, analysis consistency, traceability, and integration with26262 existing safety engineering practices.

These enhancements include the development of a detailed methodological guide, the introduction of standardised Psy-STPA worksheets, the formalisation of end-to-end traceability mechanisms, refinements to PsySIL assessment guidance, and alignment guidance with established automotive safety standards such as ISO 26262 and SOTIF. Together, these additions transform AV-PsySafe from a primarily conceptual framework into a practical engineering methodology suitable for deployment and evaluation in realistic system development contexts.

### 3.2.1. Risk Model Enhancements

The proposed psychological risk model adds enhancements to the previous AV-PsySafe risk model, by refining the psychological stake into the previously described four psychological components which are trust, perceived safety, predictability, and perceived support (Sirgabsou et al., 2025). See Figure 1.

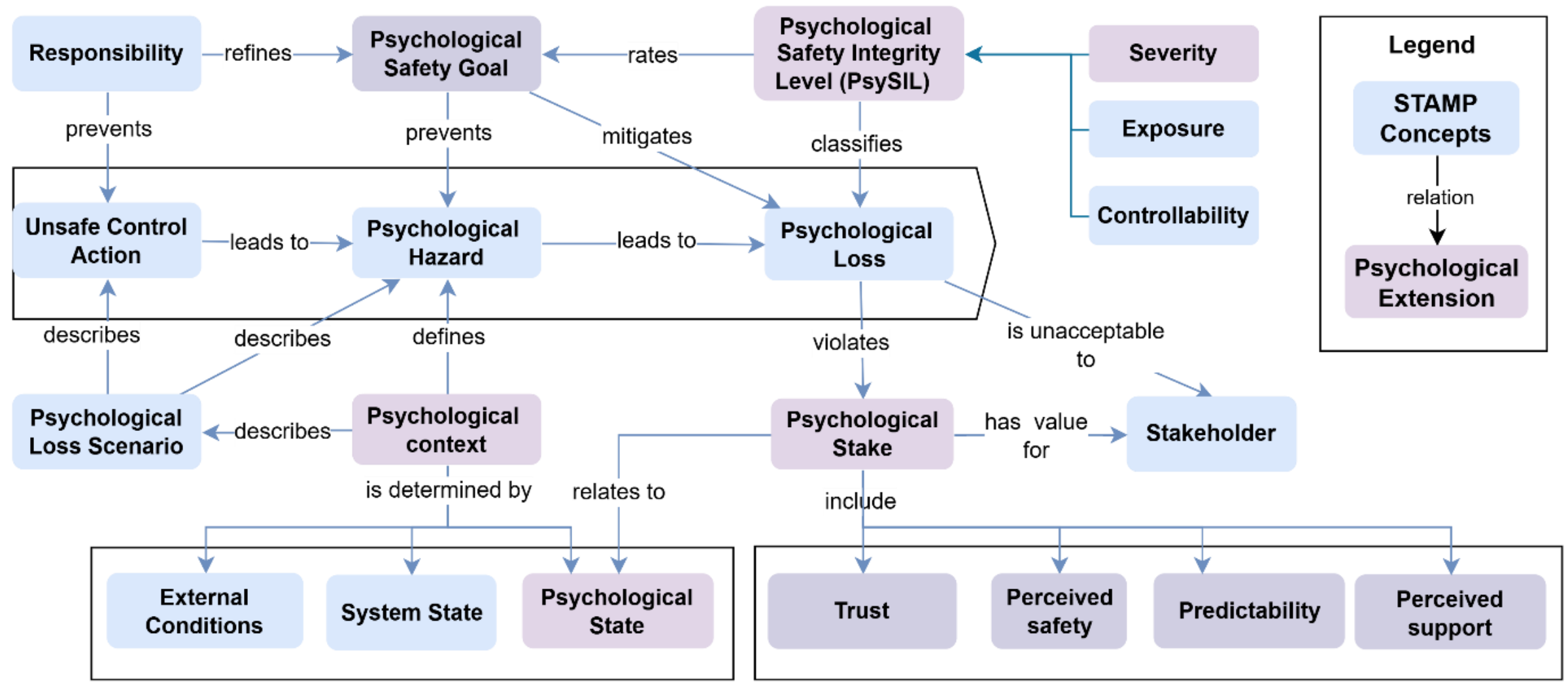

*Figure 1. Updated Psychological Safety Risk Model Based on* (*Sirgabsou et al., 2025*).

Additionally, the extended model adds traceability relations to the STPA artefacts (including responsibility, unsafe control actions, and psychological loss scenarios).

Finally, the new risk model introduces an element of psychological context which is key in both defining psychological hazards and describing psychological loss scenarios (i.e. how psychological losses occur).

### 3.2.2. Methodological Enhancements

To support consistent application of the framework, a detailed methodological guide was developed specifying the objectives, inputs, activities, and expected outputs associated with each Psy-STPA step. In addition, a set of standardised Psy-STPA worksheets was introduced to support the capture of analysis artefacts and improve consistency between analysts. These artefacts provide explicit support for documentation, traceability, and review activities, thereby improving repeatability and facilitating practical adoption of the methodology in engineering projects.

#### 3.2.3. PsySIL Assessment Guidance

The Psychological Safety Integrity Level (PsySIL) was introduced in previous work as a risk prioritisation mechanism inspired by the Automotive Safety Integrity Level (ASIL) concept used in ISO 26262. Table 1 illustrates the determination of PsySILs, ranging from A to D based on three parameters: Severity, Exposure and Controllability. Alternatively, if the risk is low, the controllability is labelled as QM, or Quality Management, meaning it is covered by normal quality standards. A detailed description of the PsySIL model, including its severity, exposure, and controllability factors, is available in the original AV-PsySafe framework (Sirgabsou et al., 2025).

*Table 1. PsySIL Assignment Matrix based on Severity, Exposure and Controllability (Sirgabsou et al., 2025).*

<table>
<tr><td colspan="3" rowspan="2"></td><td rowspan="2">Exposure Class</td><td colspan="3">Controllability class</td></tr>
<tr><td>C1<br>Simply Controllable</td><td>C2<br>Normally controllable</td><td>C3<br>Difficult to control</td></tr>
<tr><td></td><td>Psychological Effect</td><td>Severity Class</td><td></td><td></td><td></td><td></td></tr>
<tr><td rowspan="12">Severity Class</td><td rowspan="4">Short Term<br>(e.g. Increased heart rate or blood pressure)</td><td rowspan="4">S1<br>Marginal</td><td>E1</td><td>QM</td><td>QM</td><td>QM</td></tr>
<tr><td>E2</td><td>QM</td><td>QM</td><td>QM</td></tr>
<tr><td>E3</td><td>QM</td><td>QM</td><td>A</td></tr>
<tr><td>E4</td><td>QM</td><td>A</td><td>B</td></tr>
<tr><td rowspan="4">Medium Term<br>(e.g. Psychological strain)</td><td rowspan="4">S2<br>Moderate</td><td>E1</td><td>QM</td><td>QM</td><td>QM</td></tr>
<tr><td>E2</td><td>QM</td><td>QM</td><td>A</td></tr>
<tr><td>E3</td><td>QM</td><td>A</td><td>B</td></tr>
<tr><td>E4</td><td>A</td><td>B</td><td>C</td></tr>
<tr><td rowspan="4">Long Term<br>(e.g. depression, anxiety)</td><td rowspan="4">S3<br>Critical</td><td>E1</td><td>QM</td><td>QM</td><td>A</td></tr>
<tr><td>E2</td><td>QM</td><td>A</td><td>B</td></tr>
<tr><td>E3</td><td>A</td><td>B</td><td>C</td></tr>
<tr><td>E4</td><td>B</td><td>C</td><td>D</td></tr>
</table>

The validation activities reported in this paper did not modify the PsySIL concept itself. Instead, they contributed to improving the consistency of its application by introducing additional guidance regarding the assessment of exposure and controllability factors, as well as explicit rationale fields supporting PsySIL justification. These refinements reduce analyst interpretation variability and strengthen traceability between psychological hazards, loss scenarios, and risk prioritisation decisions.

#### 3.2.4. Traceability and Integration Principles

AV-PsySafe operates on operational scenarios already established through system engineering and design activities. Consequently, the methodology focuses on identifying and analysing psychological risks arising within realistic operational situations rather than defining or validating behavioural scenarios.

A key characteristic of the framework is its support for end-to-end traceability between psychological safety artefacts. The resulting analysis establishes explicit links between psychological stakes → losses → hazards → safety goals → responsibilities → UCAs → loss scenario.

This traceability enables psychological concerns to be progressively transformed into actionable engineering artefacts supporting risk assessment, mitigation, and validation activities. Furthermore, alignment guidance was introduced to facilitate integration of AV-PsySafe outputs with existing automotive safety engineering processes, particularly those associated with ISO 26262 and SOTIF.

# 4. Framework Validation in an Autonomous Driving Use Case

## 4.1. Validation strategy

This paper follows a methodological validation strategy to assess the applicability, usability, and added value of the previously proposed AV-PsySafe psychological safety framework. Rather than validating the framework through experimental user studies, the Framework Authors (FA) proposed that the validation is conducted by an independent Validation Team. The VT was identified from their contribution on real-world autonomous vehicle (AV) project and extensive experience in physical safety engineering.

The core hypothesis underlying this method is that a framework is valid for engineering practice if independent analysts are able to apply it consistently, produce meaningful results, and perceive clear value in combination with existing physical safety analyses.

In this context the FA designed a validation kit containing a guide, a template and a feedback form that was subsequently passed to the validation team to be applied on an operational use case. This application was designed to be iterative, allowing frequent checks between the VT and FA throughout the process in order to capture necessary adjustments as early as possible and provide necessary guidance if needed.

In practice, the VT was expected to:

1) apply the framework on three operational scenarios described later, relying on the validation kit,
2) systematically record their results on the provided template,
3) and provide their feedback on the framework through the provided feedback form.

Following the application of the framework, the VT's technical results and feedback were systematically collected and analysed by the FA to determine the validity of the framework. It is important to remind that the objective of the validation is not to validate a particular vehicle design, but to validate the AV-PsySafe framework itself.

Going from the two general concerns mentioning earlier in section 1 (introduction), the validation therefore focused on the following four research questions:

- RQ1 – Can AV-PsySafe be effectively applied by independent analysts?

- RQ2 – Does AV-PsySafe generate useful psychological safety artefacts and engineering insights?
- RQ3 – Is AV-PsySafe suitable for integration into industrial safety engineering processes?
- RQ4 – Is the methodology sufficiently robust to support refinement and continuous improvement?

To answer these questions, the validation relied on the present strategy which is based on two complementary sources of evidence (technical results and structured feedback). The validity of the framework was assessed after the evaluation of the collected evidence from the framework application on the previously mentioned autonomous vehicle project use case, supported by a validation kit.

## 4.2. AUTONOMOUS DRIVING USECASE CONTEXT

The validation was conducted through the application of the AV-PsySafe framework within the Route 25 autonomous vehicle project (Route 25, n.d.). Route 25 is a Portuguese autonomous mobility initiative focused on the development of safe and sustainable autonomous driving technologies among 25 national partners.

### 4.2.1. Use Case Description

The framework was applied to the *Assisted and Autonomous Driving for Safe Mobility* work package of the Route 25 project, which focuses on Advanced Driver Assistance Systems (ADAS) and Autonomous Driving (AD). The selected system of interest was the Safety Monitor module of a Level 3 autonomous driving system operating in an Urban Driving Mode (UDM) compliant with SAE J3016. The Safety Monitor is responsible for supervising vehicle dynamics and enforcing safety constraints by monitoring parameters such as vehicle speed, steering angle, and steering angle rate. When predefined safety thresholds are exceeded, the module initiates corrective actions including ADS disengagement and driver notifications.

### 4.2.2. Operational Scenarios

The validation was performed using three representative operational scenarios involving Safety Monitor interventions: overspeeding, oversteering, and oversteering rate. These three scenarios were derived from real world test scenario involving a real vehicle in a restricted test area. The tests were conducted in daylight, good weather, on multi-lane urban roads, with no traffic, pedestrians, cyclists, or parked vehicles present, and only predefined obstacles were introduced as part of the test scenarios. The Safety Monitor is expected to detect the violation of a predefined operational threshold and responds by disengaging the ADS and notifying the safety driver.

*Table 2 Operational scenarios used for the validation of AV-PsySafe within the Route 25 Safety Monitor use case*

| Scenario | Monitored Variable | Trigger Condition | Safety Monitor Response |
|---|---|---|---|
| Overspeeding | Vehicle speed | Speed > 40 km/h | ADS disengagement and driver warning |
| Oversteering | Steering angle + speed | \|angle\| > 175° and speed > 25.2 km/h | ADS disengagement and driver warning |

| Oversteering Rate | Steering rate + speed | $\lvert \text{rate} \rvert > 175°/\text{s}$ and speed > 25.2 km/h | ADS disengagement and driver warning |
|---|---|---|---|

The overspeeding scenario addresses excessive vehicle speed during autonomous operation. The oversteering scenario captures situations where large steering wheel angles indicate potentially unsafe vehicle manoeuvres, while the oversteering-rate scenario addresses abrupt steering inputs and rapid corrective actions. Together, these scenarios provide representative examples of Safety Monitor interventions involving different vehicle dynamic parameters while sharing a common mitigation strategy based on ADS disengagement and driver takeover. All validation results presented in this paper were obtained through the application of AV-PsySafe within the context of these three operational scenarios.

## 4.3. VALIDATION KIT

The validation relies on a dedicated validation kit composed of three complementary elements. First, a methodological guide provides step-by-step instructions for applying the AV-PsySafe framework within an autonomous vehicle development context (See Annex 1). Second, a set of standardized analysis templates, implemented as Psy-STPA worksheets, supports the systematic capture, documentation, and traceability of analysis artefacts and expected framework deliverables (See Annex 2). Third, a structured feedback form is used to collect both qualitative and quantitative observations from the Validation Team regarding the framework's clarity, usability, perceived usefulness, and potential integration with existing physical safety activities (Annex 1).

Together, these artefacts support consistent framework application while providing the complementary evidence required for validation.

### 4.3.1. Methodological Guide

The methodological guide acts as the main document for validation. It describes step-by-step how to apply the AV-PsySafe framework in an AV development context. The methodological guide provides:

- Definitions of psychological safety concepts,
- Guidance on how psychological safety aligns with ISO 26262 and SOTIF artifacts,
- Step-by-step instructions for each Psy-STPA phase,
- Expected inputs, activities, and outputs for each step,
- Examples illustrating typical psychological losses, hazards, UCAs, and scenarios.

The VT is instructed to strictly follow the guide, ensuring that observed outcomes directly reflect the framework's clarity and structure and to support them in the application of the framework.

### 4.3.2. Analysis Templates (Psy-STPA Worksheets)

The Psy-STPA worksheet is a standardized analysis templates to structure, document, and trace the psychological safety analysis results. It includes:

- System definition and stakeholder matrix,
- Psychological stakes and losses tables,
- Psychological hazard identification sheets,
- Psychological safety goals and responsibility allocation tables,
- Psychological control structure descriptions,
- Unsafe Control Action (UCA) identification matrices,
- Psychological loss scenario descriptions and PsySIL assignments and
- Traceability matrixes

To ensure consistency and traceability, the validation team was asked to document their work using the provided Psy-STPA Worksheets. These templates serve two complementary validation objectives: improving repeatability by constraining analyst interpretation and providing tangible evidence of framework outputs for subsequent review and evaluation.

#### 4.3.3. Structured Feedback Form

To complement the technical evidence collected through the Psy-STPA Worksheets, a structured feedback form was developed to assess the validity, usability, and practical relevance of the AV-PsySafe framework from the perspective of the analysts applying it. The form captures both quantitative ratings and qualitative insights across five dimensions:

- Background Information
- Clarity and Learnability (5-point Likert scale: strongly disagree → strongly agree)
- Usability and Effort
- Perceived Value
- Qualitative Feedback (Open-ended)

The collected responses were subsequently analysed as a complementary source of validation evidence alongside the technical artefacts produced during framework application.

### 4.4. VALIDATION PROCESS

#### 4.4.1. Iterative framework application

The VT, constituted by senior engineers from the Route 25 project, with background in physical safety analysis, were the team who conducted the project's original physical safety analysis on this operational use case. The team was asked to apply the AV-PsySafe Framework on the three selected operational scenarios. None of the team members had prior experience using a STAMP framework. The validation was conducted through three iterative application cycles of the AV-PsySafe framework. During these iterations, the VT performed the analyses using the methodological guide and Psy-STPA templates, while intermediate results and observations were periodically reviewed by the FA. Regular review meetings facilitated the discussion of findings, clarification of methodological questions, and identification of framework improvement opportunities. The resulting refinements were incorporated into subsequent analysis iterations and recorded as additional validation evidence. Consequently, the validation process served not only to evaluate the framework's applicability and usability but also to assess its robustness and ability to evolve through practical application.

#### 4.4.2. Validation evidence collection

Following completion of the validation process, two complementary categories of evidence were collected. The first category consisted of technical artefacts generated during the Psy-STPA analysis, including psychological losses, hazards, psychological safety goals, responsibilities, UCAs, loss scenarios, PsySIL assignments, and traceability matrices. The second category consisted of structured feedback provided by the Validation Team regarding conceptual clarity, usability, perceived value, effort, and industrial applicability. These two evidence sources were subsequently analysed independently and compared to establish validation conclusions.

#### 4.4.3. Validation evidence analysis

The results were analysed and used to conclude on the applicability of the framework in a real-world industrial context as well as the identification of potential improvement areas. First, the technical outputs produced during framework application by the VT are collected and analysed by the FAs to assess completeness, consistency, and engineering usefulness. Second, structured feedback is collected from the VT and evaluated to assess clarity, usability, perceived value, and industrial applicability. The convergence of these independent sources of evidence is then used to validate the framework's applicability, utility, industrial relevance, and methodological robustness.

# 5. RESULTS

In accordance with the validation strategy described in Section 4, two complementary sources of evidence were collected during the application of AV-PsySafe: (i) the technical artefacts generated through the Psy-STPA worksheets and (ii) structured feedback provided by the Validation Team (VT). These sources were first analysed independently and subsequently compared to establish validation conclusions.

## 5.1. ANALYSIS OF TECHNICAL RESULTS (PSY-STPA WORKSHEET)

#### 5.1.1. Completeness of framework application

The Validation Team successfully completed all phases of the AV-PsySafe methodology, including preparation, psychological losses and hazards dentification, control structure modelling, Unsafe Control Action (UCA) identification, and psychological loss scenario analysis. All expected artefacts were produced, demonstrating the ability of the framework to be applied end-to-end within a realistic autonomous driving use case. The successful completion of all methodological steps is particularly significant given that the analysis was performed by an independent Validation Team with no prior experience in psychological safety assessment.

Table 2 summarizes the artefacts generated during the application of the framework.

*Table 3. Generated AV-PsySafe artefacts*

| Artefact Type | Number Generated | Representative Output |
|---|---|---|
| Psychological Losses | 6 | L-01: Loss of trust in ADS after unexpected disengagement |
| Psychological Hazards | 9 | H-01: Sudden ADS disengagement due to speed violation |
| Psychological Safety Goals | 7 | PSG-01: System shall inform the driver before ADS disengagement |
| Responsibilities | 7 | R-03: Notify driver of intervention |
| Unsafe Control Actions (UCAs) | 14 | UCA-01: Disengagement without warning |
| Psychological Loss Scenarios | 14 | LS-01: Sudden ADS disengagement causes surprise and stress |

Furthermore, complete traceability was established between psychological stakes, losses, hazards, safety goals, responsibilities, UCAs, and psychological loss scenarios. This demonstrates the capacity of the framework to transform psychological concerns into structured engineering artefacts suitable for systematic analysis and mitigation.

#### 5.1.2. Engineering Value of Identified Results

Beyond the generation of artefacts, the analysis identified psychological risks that are not typically addressed through conventional safety analyses. These include risks associated with loss of trust, reduced perceived control, inadequate system feedback, insufficient explanations, cognitive overload, and unexpected vehicle behaviour.

The methodology further enabled the derivation of actionable psychological safety goals intended to mitigate these risks. Examples include providing anticipatory notifications prior to ADS disengagement, explaining intervention causes in understandable language, supporting driver situational awareness during takeover situations, prioritising alerts to reduce cognitive overload, and providing completion notifications following interventions.

*Table 4. Examples of identified psychological risks and derived psychological safety goals*

| Identified Hazard/Losses | Associated Psychological Impact | Derived Safety Goal / Mitigation |
|---|---|---|
| Sudden ADS disengagement due to speed violation (H-01) | Loss of trust, stress, reduced predictability | PSG-01: Inform the driver before ADS disengagement |
| Unexpected switch between manual and ADS (H-06) | Confusion, loss of perceived control | PSG-04: Support driver situational awareness during takeover |
| No feedback after intervention completion (H-08) | Feeling abandoned by automation, long-term trust degradation | PSG-05: Confirm intervention completion with contextual notification |
| Multiple simultaneous interventions (H-09) | Cognitive overload and situational awareness loss | PSG-06: Prioritise and sequence alerts |
| Emergency braking without visible cause (LS-04) | Reduced trust and perceived control | Explain braking causes through contextual feedback |

The generated artefacts reveal psychological risks that are generally not captured by conventional safety analyses. Notably, the framework identified hazards associated with unexpected mode transitions, post-intervention silence, cognitive overload caused by concurrent alerts, and trust degradation following unexplained interventions. These risks were systematically translated into actionable safety goals related to anticipation, explanation, situational awareness, feedback reliability, and cognitive load management.

#### 5.1.3. Evidence of Methodological Maturity

The validation exercise also provided evidence of methodological maturity. Across the three application cycles, the Validation Team identified several opportunities to improve the practical application of the framework. These included the clarification of PsySIL assessment criteria, particularly regarding the interpretation of exposure and controllability factors, the strengthening of traceability mechanisms between generated artefacts, and the introduction of explicit rationale fields to support risk assessment decisions. In addition, improvements to the analysis templates enhanced the consistency and usability of the methodology.

All identified refinements were incorporated during subsequent iterations without modifying the underlying concepts, process structure, or risk model of AV-PsySafe. This ability to accommodate analyst-driven improvements while preserving methodological coherence provides evidence of both framework robustness and capacity for continuous evolution. Such adaptability is particularly important for emerging engineering approaches, which are expected to mature through iterative application and feedback from practitioners.

### 5.2. Analysis of Structured Feedback

The structured feedback collected from the Validation Team (VT) provided complementary evidence regarding the clarity, usability, perceived value, and industrial applicability of the AV-PsySafe framework. The feedback was gathered following completion of the analysis activities and was used to assess the framework from the perspective of practitioners applying the methodology in a realistic engineering context. Table 5 provides an overview of the feedback across criteria.

#### 5.2.1. Conceptual clarity

The VT strongly agreed that the principal concepts underpinning the framework, including psychological safety, psychological losses, psychological hazards, Unsafe Control Actions (UCAs), and Psychological Safety Integrity Levels (PsySIL), were understandable and sufficiently defined for engineering applications. Importantly, the analysts reported no prior experience with psychological safety assessment. This indicates that the AV-PsySafe framework can be learned and applied by practitioners without requiring specialised psychological safety expertise.

#### 5.2.2. Usability and practicality

The VT considered the methodological guide sufficient for conducting the analysis and highlighted the usefulness of the Psy-STPA templates in structuring activities and maintaining traceability between artefacts. The framework was further assessed as realistic for industrial use, with the required effort and associated cognitive workload considered acceptable for engineering projects.

#### 5.2.3. Perceived added value

The analysts reported that the methodology introduced a complementary perspective to traditional safety analyses by explicitly addressing psychological risks arising from human–automation interactions. The framework was perceived as capable of identifying actionable mitigation measures aimed at improving trust, confidence, predictability, and overall system acceptability. The Validation Team also highlighted the value of considering psychological and physical safety concerns together when analysing autonomous driving systems.

#### 5.2.4. Industrial relevance

The VT assessed the framework as highly compatible with established automotive safety practices and identified strong integration potential with ISO 26262 and SOTIF activities. Rather than replacing existing safety analyses, the framework was consistently perceived as a complementary method extending current safety engineering practices towards psychological risk assessment.

*Table 5. Validation Team feedback on clarity, usability, and integration potential*

| Criterion | Assessment |
|---|---|
| Conceptual clarity | Strong agreement |
| Usability | Strong agreement |
| Industrial applicability | Strong agreement |
| ISO26262 compatibility | 5/5 |
| SOTIF compatibility | 5/5 |

Overall, the feedback results indicate that the framework is understandable, usable, and relevant to industrial safety engineering practice

### 5.3. Cross-Analysis of Complementary Evidence

The validation strategy relied on two complementary sources of evidence: (i) the technical artefacts generated during framework application and (ii) the structured feedback provided by the Validation Team. These evidence sources were analysed independently and then compared to answer the four research questions defined in Section 4.

#### 5.3.1. Validation of Applicability (RQ1)

The technical analysis demonstrated that independent analysts successfully completed all phases of the methodology and generated the expected psychological safety artefacts. Concurrently, feedback results showed that the framework was understandable, learnable, and usable despite the analysts having no prior experience in psychological safety. The convergence of these findings provides strong evidence that AV-PsySafe can be effectively applied in realistic engineering contexts.

#### 5.3.2. Validation of Utility or Usefulness (RQ2)

The generated artefacts demonstrate the framework's ability to identify psychological risks that are not typically addressed in conventional safety analyses and to derive corresponding psychological safety goals and mitigation measures. Feedback further confirmed that these outputs were perceived as valuable, actionable, and complementary to existing safety engineering practices. Together, these results validate the practical utility of the framework.

#### 5.3.3. Validation of Industrial Relevance (RQ3)

The structured nature of the generated artefacts, combined with the use of traceability mechanisms and PsySIL-based risk prioritisation, provides objective evidence of engineering compatibility. This is reinforced by feedback indicating strong alignment with ISO 26262 and SOTIF practices and a high potential for industrial adoption. The convergence of these findings validates the industrial relevance of AV-PsySafe.

#### 5.3.4. Validation of Methodological Robustness (RQ4)

Both sources of evidence indicate that the framework is capable of supporting methodological refinement without compromising its conceptual foundations. Technical results demonstrate that several improvements were identified and successfully incorporated during the validation process, while feedback results highlighted additional opportunities for future enhancement without questioning the overall usefulness of the methodology. These findings provide evidence that AV-PsySafe possesses sufficient robustness and maturity to support continuous improvement.

*Table 6. Cross-analysis of technical and feedback evidence against the validation research questions*

| Research Question | Technical Evidence | Feedback Evidence | Conclusion |
|---|---|---|---|
| RQ1 – Can AV-PsySafe be effectively applied by independent analysts? | Complete execution of all Psy-STPA steps and generation of expected artefacts | Strong ratings for clarity, learnability, and usability | Validated |
| RQ2 – Does AV-PsySafe generate useful psychological safety artefacts and engineering insights? | Identification of psychological risks, safety goals, UCAs, and loss scenarios | Outputs perceived as useful and actionable | Validated |
| RQ3 – Is AV-PsySafe suitable for integration into industrial safety engineering processes? | Traceability mechanisms and PsySIL-based prioritisation | Strong compatibility with ISO 26262 and SOTIF | Validated |
| RQ4 – Is the methodology sufficiently robust to support refinement and continuous improvement? | Successful integration of improvements during validation | Positive assessment despite identified improvement opportunities | Validated |

The convergence between technical results and analyst feedback provides consistent evidence supporting the applicability, utility, industrial relevance, and robustness of the AV-PsySafe framework. Collectively, these findings indicate that the methodology constitutes a viable and engineering-oriented approach for the systematic assessment of psychological safety in autonomous vehicle systems.

## 5.4. Challenges Identified During Validation

Although the Validation Team successfully completed the methodology and produced all expected artefacts, the iterative validation exercise identified three main areas for improvement. These observations primarily concern methodological guidance and scalability rather than the underlying principles of the AV-PsySafe framework.

First, the VT highlighted the need for additional guidance to support the construction of the psychological safety control structure. While the control structure was successfully developed, the validation process revealed that both the methodological guide and analysis template could de improved for better consistency and reduced analysis effort. In particular, the STAMP graphical control structures are built around the concept of hierarchical control (Leveson, 2012), a perspective that is not commonly used in traditional systems engineering practices. As a result, the hierarchical relationships between controllers and controlled processes were not immediately intuitive to the VT. Consequently, more explicit instructions, graphical examples, and best-practice recommendations were added in the methodological guide and Psy-STPA template by the FAs. Furthermore , it would benefit the framework if structured details on hierarchical levels in STAMP control structures were provided, specially for people with little experience with STAMP.

Second, the validation process identified opportunities to make the formulation of analysis artefacts even more standardized, particularly control actions, Unsafe Control Actions (UCAs), and psychological loss scenarios. Although the generated artefacts were considered relevant and complete, the validation process revealed that their formulation could be improved to ensure consistency across analyses and facilitate future automation and tooling support. Consequently, additional formulation guidance, naming conventions, and boilerplate structures were added to the methodological guide (e.g. see boilerplates, best practices in Annex 1).

Third, maintaining traceability between psychological losses, hazards, safety goals, responsibilities, UCAs, and loss scenarios became increasingly demanding as the analysis evolved. While complete traceability was successfully achieved, the VT identified the need for stronger traceability support mechanisms and clearer derivation rules to improve scalability and ease of application in larger and more complex industrial projects. While the work-around consisted in a thorough manual traceability consistency check, better traceability could be supported by a tool, such as macros, requirement-management tool, or model-based environments.

Importantly, none of these challenges prevented successful application of the framework. Rather than revealing limitations of the AV-PsySafe concepts themselves, they highlight opportunities to further improve methodological guidance, consistency, and scalability. Several of the resulting recommendations were incorporated during the validation process, contributing to the refinement of the framework and providing additional evidence of its capacity for continuous improvement.

### 5.5. Summary of Validation Evidence

The combined analysis of technical artefacts and structured feedback provides consistent and convergent evidence supporting the applicability, utility, industrial relevance, and robustness of the AV-PsySafe framework. The technical results demonstrate that independent analysts were able to successfully execute the complete methodology and generate a comprehensive and traceable set of psychological safety artefacts. The feedback results further confirm that the framework is understandable, usable, and relevant for engineering practice despite the analysts having no prior experience in psychological safety assessment.

The validation additionally demonstrated the framework's ability to identify psychological risks that are generally not addressed by conventional safety analyses and to translate them into actionable psychological safety goals and mitigation measures. Furthermore, the positive assessment of its compatibility with ISO 26262 and SOTIF activities indicates strong potential for integration within existing automotive safety engineering processes.

Beyond the production of analysis artefacts, the validation exercise contributed directly to the maturation of the framework. Successive application cycles led to improvements in PsySIL assessment support, traceability mechanisms, and methodological guidance, which were incorporated without modification of the framework's underlying concepts. For instance, more detailed explanations on PsySIL determination were added to the Psy-STPA template to guide VT. Additionally, further guidance and on control structure building were added to both the PsySafe guide and the Psy-STPA worksheet. Moreover, boilerplate for better Psy-STPA artefacts formulation were added the guide. Finaly, the traceability tables were improved thanks to the VT's feedback and suggestions during the iterative process. This demonstrates not only the practical applicability of AV-PsySafe but also its capacity for continuous refinement through operational use and practitioner feedback.

Overall, the validation results provide positive answers to all four research questions and indicate that AV-PsySafe constitutes an engineering-ready methodology for the systematic assessment of psychological safety in autonomous vehicle systems.

# 6. Discussion

## 6.1 Psychological Safety as an Engineering Concern

A central outcome of this work is the demonstration that psychological safety can be considered through a structured engineering methodology rather than solely through behavioural studies, user evaluations, or human factors assessments. The successful application of AV-PsySafe by an independent Validation Team with no prior experience in psychological safety indicates that psychological losses, hazards, safety goals, and mitigation measures can be systematically identified through engineering artefacts and processes.

This finding extends previous research on trust in automation and human–automation interaction. Trust, perceived control, predictability, and situational awareness have long been recognised as fundamental determinants of effective interaction with automated systems (Lee and See, 2004; Endsley, 1995). Similarly, studies on autonomous vehicle acceptance have highlighted that user adoption depends not only on objective safety but also on subjective perceptions of trustworthiness and transparency (Hegner et al., 2019; Shariff et al., 2017). However, these concepts have traditionally been studied as behavioural constructs rather than as elements of a structured safety engineering process. The present work contributes to making these concepts operational within a systems-theoretic hazard analysis framework and linking them directly to engineering decisions and mitigation measures.

## 6.2 Added Value Beyond Traditional Safety Analyses

The validation results indicate that AV-PsySafe augments rather than replaces existing safety engineering approaches. Traditional automotive safety analyses, including ISO 26262 and SOTIF, focus primarily on preventing physical harm arising from failures or functional performance limitations. While these approaches are essential, they do not explicitly address psychological losses resulting from human–automation interactions. The present analysis identified several categories of psychological risk, including trust degradation following unexplained interventions, loss of perceived control during sudden system actions, uncertainty caused by ambiguous mode transitions, and cognitive overload resulting from multiple concurrent alerts—that would not typically be captured through conventional safety assessments.

These findings are consistent with previous studies showing that concerns related to uncertainty, transparency, and inadequate explanations constitute important barriers to autonomous vehicle adoption (Shariff et al., 2017; Yoo et al., 2025). The framework further demonstrated the ability to transform these concerns into actionable psychological safety goals such as anticipatory communication, intervention explanations, takeover support, feedback mechanisms, and alert prioritization. Consequently, AV-PsySafe extends established safety analyses by introducing a complementary perspective centred on the psychological consequences of system behaviour while remaining compatible with existing safety engineering objectives.

## 6.3 Industrial Applicability and Integration Potential

The validation provides encouraging evidence regarding the applicability of AV-PsySafe in industrial settings, such as the Route 25 consortium's scenarios. The methodology was consistently assessed as understandable, usable, and realistic for engineering deployment, while its artefacts were perceived as compatible with existing development processes. The Validation Team further identified strong integration potential with ISO 26262 and SOTIF activities, supporting the view that psychological safety assessment can be incorporated into established automotive safety workflows.

This result aligns with the systems-theoretic perspective proposed by Leveson (2012), according to which safety emerges from interactions between technical, organisational, and human elements rather than from component reliability alone. By preserving familiar concepts such as hazard identification, risk prioritisation, traceability, and safety constraints, AV-PsySafe facilitates integration with existing engineering practices while extending them to address psychological concerns. The introduction of structured artefacts, responsibilities, and PsySIL-based prioritisation further strengthens this compatibility by providing outputs that are readily understandable to safety practitioners.

## 6.4 Methodological Maturity and Continuous Improvement

An important contribution of the validation concerns the framework's ability to evolve through practical application. Throughout the three validation cycles, analysts identified opportunities to strengthen PsySIL assessment guidance, artefact traceability, and methodological support. These refinements resulted in improvements to the templates, traceability mechanisms, and assessment guidance without requiring modifications to the underlying risk model or Psy-STPA process.

This behaviour is characteristic of mature engineering methodologies, which evolve through repeated deployment and practitioner feedback rather than solely through theoretical development. Similar observations have been reported during the evolution of STAMP and STPA, where practical application has progressively refined methodological guidance while preserving the underlying systems-theoretic foundations (Jimeno Altelarrea et al., 2022; N. G. Leveson, 2012; N. Leveson & Thomas, 2018). Rather than exposing weaknesses in AV-PsySafe, the identified improvements demonstrate conceptual stability combined with adaptability, supporting its long-term viability as an engineering methodology.

## 6.5 Limitations and Future Work

Several limitations should nevertheless be acknowledged. First, the validation was conducted on a single subsystem within the Route 25 autonomous driving project. Although the use case provided realistic operational scenarios and enabled complete application of the methodology, additional studies involving different vehicle functions, automation levels, and industrial environments could contribute to the generalization of the findings.

Second, the present work should be viewed as a validation of the methodology rather than a validation of psychological outcomes themselves. While the framework successfully identified psychological hazards and derived mitigation measures, future studies should investigate the relationship between these outputs and measurable human factors constructs such as trust, workload, perceived control, situational awareness, and system acceptance. Such work would establish stronger empirical links between the identified psychological risks and observed user responses. This direction is consistent with the broader literature on trust in automation and situation awareness (Endsley, 1995; Lee & See, 2004).

Finally, the validation highlighted opportunities for further methodological enhancement, particularly regarding control structure modelling, artefact formulation, and traceability support. Dedicated tooling, improved automation support, and additional guidance could further improve scalability and facilitate deployment in large industrial projects.

## 6.6 Towards the Co-Assessment of Physical and Psychological Safety

Although the objective of this work was not to establish a unified physical and psychological safety assessment process, the validation results suggest that such an evolution is feasible. The strong compatibility observed between AV-PsySafe artefacts and existing automotive safety practices indicates that psychological safety can be incorporated within established engineering workflows without disrupting existing processes.

This perspective aligns with the systems-theoretic view that safety emerges from interactions between human and technical elements and therefore cannot be fully understood through physical hazard analysis alone (Leveson, 2012). Future research should investigate mechanisms for linking physical hazards, psychological hazards, integrity levels, safety requirements, and mitigation measures within a common systems engineering framework.

Such developments would support a more comprehensive assessment of autonomous vehicle safety and user acceptability, contributing to the design of systems that are not only physically safe but also trustworthy, understandable, and psychologically acceptable

# 7. CONCLUSION

Psychological factors such as trust, perceived control, predictability, and perceived support play an important role in the acceptance and effective use of autonomous vehicle systems. However, despite extensive research in human factors and behavioural sciences, these concerns remain insufficiently addressed within established safety engineering practices. This work addressed this gap through the refinement and validation of AV-PsySafe, a systems-theoretic framework for the systematic assessment of psychological safety risks in autonomous vehicles.

Building upon the AV-PsySafe framework first version, this work introduced a set of methodological enhancements aimed at supporting its practical application in engineering contexts. This includes a detailed methodological guide, standardized Psy-STPA templates, strengthened traceability mechanisms, refined PsySIL assessment guidance, and alignment considerations with existing automotive safety processes. These improvements transformed the framework from a conceptual proposal into a practical engineering methodology.

The framework was subsequently validated through its application to a realistic autonomous driving use case within the Route 25 project by an independent Validation Team. The results demonstrated that AV-PsySafe can be successfully applied by analysts without prior expertise in psychological safety. They show that AV-PsySafe can help in generating complete and traceable psychological safety artefacts, identifying psychological risks not typically captured by conventional safety analyses and supporting the derivation of actionable mitigation measures. The validation further provided evidence of the framework's applicability, utility, industrial relevance, and robustness while identifying opportunities for continued methodological improvement. The updates to the AV-PsySafe framework are broadly applicable across scenarios where it may be used, and we encourage designers to use this updated version of the framework.

Overall, the findings indicate that AV-PsySafe constitutes an engineering-ready methodology for the systematic identification, analysis, prioritisation, and mitigation of psychological risks in autonomous vehicle systems. Beyond autonomous driving, the proposed approach contributes to the broader objective of integrating human-centred considerations into systems and safety engineering processes. Future work will focus on expanding validation across additional industrial contexts, strengthening integration with established safety standards, investigating relationships between identified psychological risks and measurable human factors outcomes, and supporting the coordinated assessment of physical and psychological safety within complex autonomous systems.

## 8. ACKNOWLEDGMENTS

The authors gratefully acknowledge the Route 25 consortium for providing the industrial use case, operational scenarios, and system context used in this validation study.

The authors would particularly like to thank the Route 25 team members, Rafael Fonte Valente, Anderson Ramos and David Ferreira, for conducting the independent validation of the AV-PsySafe framework.

## 9. REFERENCES

Please insert all references before this line